\documentclass[10pt,twocolumn,letterpaper]{article}
\usepackage{authblk}
\usepackage[pagenumbers]{wacv_arxiv}

\usepackage{graphicx}
\usepackage{amsmath}
\usepackage{amssymb}
\usepackage{booktabs}
\usepackage{multirow}
\usepackage[accsupp]{axessibility}

\usepackage[pagebackref,breaklinks,colorlinks]{hyperref}

\usepackage[capitalize]{cleveref}
\crefname{section}{Sec.}{Secs.}
\Crefname{section}{Section}{Sections}
\Crefname{table}{Table}{Tables}
\crefname{table}{Tab.}{Tabs.}

\usepackage{caption}

\newcommand{\beginsupplement}{
    \setcounter{section}{0}
    \renewcommand{\thesection}{S\arabic{section}}
    \setcounter{figure}{0}
    \renewcommand{\thefigure}{S\arabic{figure}}
    \setcounter{table}{0}
    \renewcommand{\thetable}{S\arabic{table}}
}

\begin{document}

\title{SoundSil-DS: Deep Denoising and Segmentation of \\Sound-field Images with Silhouettes\vspace{-12pt}}

\author[1,2]{Risako Tanigawa}
\author[1]{Kenji Ishikawa}
\author[1]{Noboru Harada}
\author[2]{Yasuhiro Oikawa\vspace{-8pt}}

\affil[1]{NTT Communication Science Laboratories, NTT Corporation, Atsugi, Kanagawa 243-0198, Japan}
\affil[2]{Intermedia Art and Science, Waseda University, Shinjuku, Tokyo 169-8555, Japan}

\affil[ ]{{\tt\small \{risako.tanigawa, ke.ishikawa, harada.noboru\}@ntt.com, yoikawa@waseda.jp}}

\maketitle

\begin{abstract}
Development of optical technology has enabled imaging of two-dimensional (2D) sound fields.
This acousto-optic sensing enables understanding of the interaction between sound and objects such as reflection and diffraction.
Moreover, it is expected to be used an advanced measurement technology for sonars in self-driving vehicles and assistive robots.
However, the low sound-pressure sensitivity of the acousto-optic sensing results in high intensity of noise on images.
Therefore, denoising is an essential task to visualize and analyze the sound fields.
In addition to denoising, segmentation of sound and object silhouette is also required to analyze interactions between them.
In this paper, we propose sound-field-images-with-object-silhouette denoising and segmentation (SoundSil-DS) that jointly perform denoising and segmentation for sound fields and object silhouettes on a visualized image.
We developed a new model based on the current state-of-the-art denoising network.
We also created a dataset to train and evaluate the proposed method through acoustic simulation.
The proposed method was evaluated using both simulated and measured data.
We confirmed that our method can applied to experimentally measured data.
These results suggest that the proposed method may improve the post-processing for sound fields, such as physical model-based three-dimensional reconstruction since it can remove unwanted noise and separate sound fields and other object silhouettes.
Our code is available at {\small \url{https://github.com/nttcslab/soundsil-ds}}.
\end{abstract}

\vspace{-10pt}
\section{Introduction}
\vspace{-2pt}
\label{sec:intro}
Sound is one of the most important cues to understanding scenes as well as vision.
For example, self-driving cars and assistive robots are equipped with ultrasonic sonars and microphones, which are used to gather information about their surroundings.
Recently, research has been conducted on converting sound information into vision information and vice versa.
Lindell \textit{et al.} have proposed an acoustic non-line-of-sight imaging method for resolving three-dimensional object shapes hidden around corners through acoustic echoes~\cite{Lindell2019ANLOS}.
Davis \textit{et al.} have proposed a method with which sound waves are recovered through object vibrations captured using a high-speed camera~\cite{Davis2014VisualMic}.
Sheinin \textit{et al.} have proposed a method of sensing sound at high speeds through object-surface vibrations~\cite{Sheinin2022DualShutter}.
These studies demonstrated the potential of capturing sound as images, paving the way for further advancements in the field.

\begin{figure}[t]
  \centering
   \includegraphics[width=1.0\linewidth]{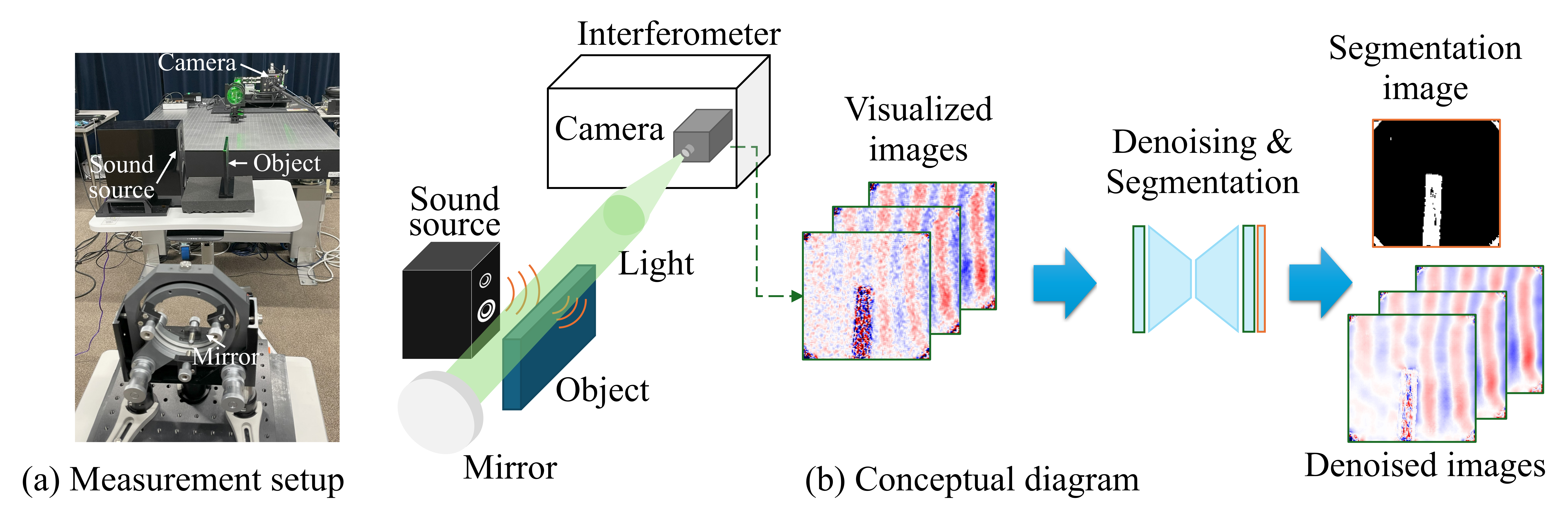}
   \caption{Conceptual diagram of proposed method. (a) Experimental setup for optical sound measurement, which is microphone-free sound measurement device. (b) Conceptual diagram. Sound field with interacting objects is captured as images with high-speed camera. Visualized images are converted to denoised and segmentation images with a DNN.}
   \vspace{-8pt}
   \label{fig:concept}
\end{figure}

A sound-visualization technique involving directly capturing and visualizing the density variations in air caused by sound has been proposed~\cite{Samuel2023}.
Such acousto-optic sensing can capture sound without any microphones by observing modulations of the phase of light passing through sound fields.
By using high-speed cameras as sensors, it becomes possible to create visual representations of invisible sound waves as images~\cite{PPSI2016,Awatsuji2021PPSDH}.
However, a significant challenge remains: the phase modulation of light induced by sound is extremely small, leading to a high level of noise in the measured images.
Ishikawa {\it et al.}~\cite{Ishikawa2023} have proposed deep sound-field denoiser (DSFD), the first denoiser to use deep neural networks (DNNs) for noise reduction in sound-field images, demonstrating that DNN-based methods achieve superior denoising performance compared with conventional filtering methods.

With the optical technology and signal processing method, we can understand the nature of sound propagation. 
Therefore, the interaction between sound and objects, such as reflections and diffractions, can be visualized using acousto-optic sensing.
To analyze such interactions between sound and objects, both denoising and segmentation of the sound field and object regions on visualized images should be done simultaneously.

We propose a method for simultaneous denoising and segmentation of sound-field images including object silhouettes.
A conceptual diagram is presented in~\cref{fig:concept}.
With this method, optically measured sound-field images with object silhouettes are denoised and segmented using a DNN.
The sound field on the laser path is captured with a high-speed camera.
The area where the object blocks the laser light can be visualized as noisy silhouettes in the visualized images.
The visualized images are converted to denoised and segmentation images with the DNN.
The DNN is constructed based on the state-of-the-art (SOTA) denoising network, which also has the potential for per-pixel feature segmentation.
We created a dataset with acoustic simulation since there is no dataset that includes acoustic scattering by objects.
Denoising and segmentation are expected to (1) enable analysis of the propagation, reflection, and diffraction of sound waves in space and (2) be used as an advanced measurement technology for sonars in self-driving vehicles and assistive robots.
The contributions of our work are summarized as follows:
\begin{itemize}
    \setlength{\itemsep}{0mm}
    \setlength{\parskip}{0mm}
    \item We propose a method for simultaneous denoising and segmentation of sound-field images with silhouettes. 
    \item We created a dataset considering acoustic scattering caused by various shapes of objects.
    \item We confirmed that the proposed method performed effectively on denoising and segmentation tasks.
\end{itemize}

\section{Related Work}
\label{sec:related}

\subsection{Acousto-optic Sensing}
Acousto-optic sensing detects sounds by illuminating the sound fields with laser light from a distance, by using the changes in the refractive index of air due to density variations caused by sound.
Since the method can capture sound without microphones, it can be useful to measure sound where microphone cannot be used such as narrow spaces and inside airflow.
This technique has been applied to various situations such as visualizing sound field generated from loudspeakers~\cite{2008Sonoda, Hargather2009, PPSI2016}, measuring flow-induced sounds~\cite{Sonoda2012JetNoise, Tanigawa2020ExpFluid}, and microphone calibrations~\cite{Koukoulas2015, Hermawanto2023}.
Using a high-speed camera to capture sound-field images provides an intuitive understanding of acoustic phenomena~\cite{PPSI2016, Tanigawa2020ExpFluid, Hermawanto2023}.

While they can non-intrusively capture 2D sound fields, the measured data often contain a significant amount of noise due to  the small modulation of the light phase.
An example of the measured data of visualized images is illustrated in~\cref{fig:concept}. 
The sound wave propagates from right to left within the images,  with reflections and diffractions occurring at the reflector. 
The data contain a significant noise; therefore, the diffracted waves are almost invisible.

\subsection{Sound-field-image Denoising}
Sound-field images captured by a high-speed camera are three-dimensional data in two dimensions of space and one dimension of time.
Sound-field-image denoising has been done for both space and time dimensions.

\paragraph{Classical Filters}
For more than a decade, classical filters have been used to reduce noise in sound fields~\cite{Yatabe2018, Chitanont2017, Tanigawa2019}.
The most straightforward method involves time-domain filtering~\cite{Yatabe2018}.
The time-directional signals in each pixel can be considered similar to microphone signals where 2D sound fields are captured by high-speed camera.
Hence, by applying time-domain filtering on a per-pixel basis, it becomes feasible to extract images corresponding to specific frequencies.
Although the time-domain filters can remove noise with frequencies other than that of sound, noise components that fluctuate at the same frequency as the sound cannot be removed.
In contrast, by using spatio-temporal filters~\cite{Chitanont2017, Tanigawa2019}, we can extract components that satisfy the frequency of sound in both the time and spatial domains.

\paragraph{Deep Sound-Field Denoiser (DSFD)\protect\cite{Ishikawa2023}}
The DNN-based sound-field denoising method DSFD has been proposed.
It is the first attempt at sound-field denoising using a DNN.
The network of DSFD is based on nonlinear activation free network (NAFNet)~\cite{NAFNet2022}, which is a network for natural image denoising.
The input data consists of frequency-domain data obtained by applying Fourier transform (FT) in the time direction to each pixel of a sound-field video.
The real and imaginary parts of the complex amplitudes at a specific frequency are treated as two separate images, which are then stacked along the channel direction, resulting in input data $X \in \mathbb{R}^{N \times C \times W \times H}$, where $N$ is the number of images, $C=2$ is the number of channels, and $W$ and $H$ are the number of pixels in width and height.
The training data has been generated from 2D sound-field simulations in the frequency domain.
Supervised learning is carried out using the simulated noisy images as inputs.
By using DNNs, denoising can be achieved with higher accuracy compared with traditional classical filtering methods.

\begin{figure*}[t]
    \centering
    \includegraphics[width=1.0\linewidth]{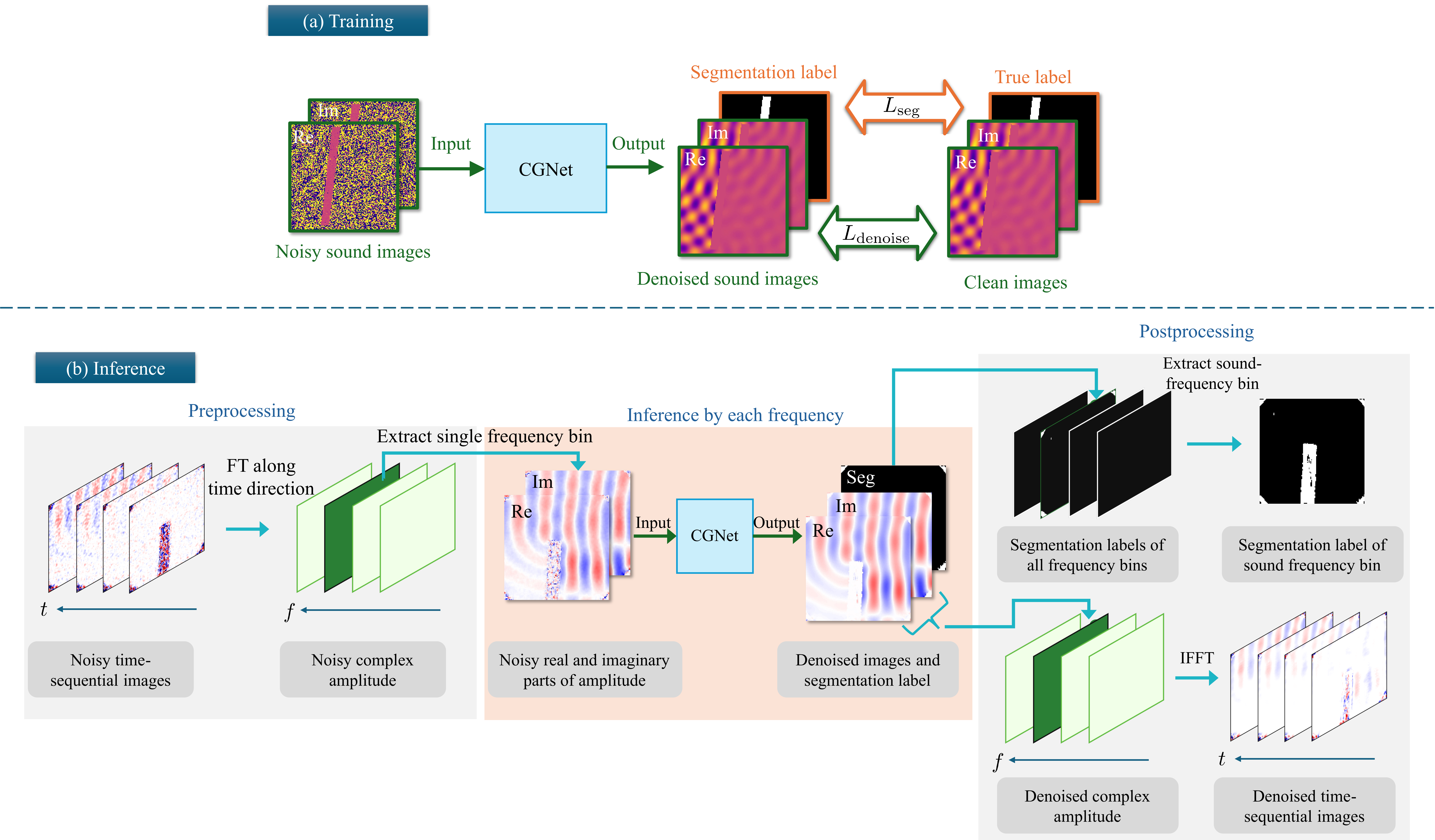}
    \caption{Overview of our approach. (a) Training process. Two channels of noisy sound images are input into network. Output images have three channels: first two channels are for denoising and last one channel is for segmentation. Loss of denoised and segmented images is calculated separately for each ground truth image. (b) Inference process. Experimentally measured time sequential images are converted to frequency domain by Fourier transform (FT). Each frequency complex amplitude is turned into real and imaginary images and input into trained model. Denoised images of all frequency bins are converted to time-sequential images by inverse FT. Segmentation image at sound frequency is extracted as final segmentation label.}
    \vspace{-6pt}
    \label{fig:overview}
\end{figure*}

\subsection{Image Denoising and Segmentation}
As demonstrated with DSFD, 2D sound-field data can be regarded as images, allowing the application of DNNs commonly used for RGB/grayscale images.
Numerous natural-image denoising models have been proposed, including convolutional neural network (CNN)-based~\cite{DnCNN, LRDUNet, NAFNet2022, KBNet, CGNet2024} and transformer-based models~\cite{Restormer2022, Uformer2022}.
Denoising models have been proposed that use lightweight CNNs instead of transformers.
In fact, the top three SOTA denoising models on the Smartphone Image Denoising Dataset (SIDD)~\cite{SIDD2018} have used CNNs~\cite{CGNet2024, KBNet, NAFNet2022}.
These CNN-based models are designed to retain global information, achieving impressive accuracy.
For instance, Cascaded Gaze Net (CGNet)~\cite{CGNet2024}, which outperforms the NAFNet~\cite{NAFNet2022}, has achieved a peak signal-to-noise ratio (PSNR) of 40.39 dB on the SIDD.
Therefore, we believed it better to use this SOTA network as a base network other than NAFNet in our study.

\vspace{-2pt}
The task of joint denoising and segmentation, which is the objective of this study, is explored in the field of microscopy imaging.
DenoiSeg~\cite{DenoiSeg2020}, for instance, is a method that performs both denoising and segmentation on biological images.
DenoiSeg achieves this multitasking by adding channels corresponding to the segmentation classes to the decoder of a denoising network in Noise2Void~\cite{Noise2Void2019}.
Losses are then calculated for each task, enabling the model to produce both denoised and segmentation images.
The largest difference of the proposed method is using self-supervised learning due to the difficulty in obtaining clear images.
Sound-field images can be generated from acoustic simulations, making supervised learning a rational approach.

\section{Proposed Methods}
\label{sec:method}
To simultaneously carry out denoising and segmentation, we use the SOTA denoising network, CGNet~\cite{CGNet2024}, as a base network.
We increase the number of output channels in the final layer of CGNet to three channels.
The first two channels are for denoising sound images, that same as with DSFD.
The last channel is for the segmentation of object silhouettes.
By increasing the number of channels in the final layer, the proposed method can accomplish both tasks, resulting in lower computational costs compared with training denoising and segmentation sequentially.

\subsection{Proposed Archtecture}
Overviews of training and inference processes are shown in~\cref{fig:overview}(a) and~\cref{fig:overview}(b), respectively.
The same as with DSFD, denoising is carried out in the frequency domain.
To achieve this, frequency-domain data, obtained by applying a 1D FT in the time direction to time-domain sound-field videos are used.
In the training process (~\cref{fig:overview}(a)), simulated sound-field images with added noise are input to CGNet.
The input channels consist of real and imaginary parts of the frequency-domain data obtained from 1D FT.
The number of channels of the last layer of the network is set to three channels: first two channels are for denoising $\hat{X}_{\rm{denoise}} \in \mathbb{R}^{N \times 2 \times W \times H}$ and last channel is for segmentation $\hat{X}_{\rm{seg}} \in \mathbb{R}^{N \times 1 \times W \times H}$.
The range of the segmentation data is converted to $\hat{X}'_{\rm{seg}} \in [0,1]$ using a sigmoid function.
Segmentation labels are the binary labels obtained by the thresholding of $\hat{X}'_{\rm{seg}}$.
The binary labels are 0 for sound and 1 for silhouette class.
The loss function is the weighted sum of $L_{\rm{denoise}}$ and $L_{\rm{seg}}$ as 
\begin{equation}
    L = L_{\rm{denoise}} + \lambda L_{\rm{seg}},
    \label{eq:lossfun}
\end{equation}
where $\lambda$ is the weighting coefficient that balances the two losses.
For denoising, the negative PSNR loss function is used as well as that of CGNet.
For segmentation, a combination of binary cross entropy loss and dice loss is used to reduce the bias in the number of pixels in the sound and silhouette classes:
\begin{equation}
    L_{\rm{seg}} = (1-\alpha)L_{\rm{BCE}} + \alpha L_{\rm{Dice}},
    \label{eq:segloss}
\end{equation}
where $\alpha$ is the weighting coefficient.
These loss functions are determined based on the result of the preliminary experiment (See details in the supplementary material).

The inference process is shown in~\cref{fig:overview}(b).
To carry out inference in the frequency domain, noisy time-sequential images $x_{\rm{raw}} \in \mathbb{R}^{W \times H \times T}$ are converted to noisy complex amplitude $X_{\rm{raw}} \in \mathbb{C}^{W \times H \times F}$ with 1D FT along the temporal axis, where $T$ and $F$ are the number of data samples in temporal and frequency axes, respectively.
Denoising and segmentation are carried out for each frequency bin of $X_{\rm{raw}}$.
The image in which the real and imaginary parts of the $i$-th frequency bin $X_{\rm{raw},\it{i}}$ are arranged in the channel direction is input to the network in the same manner as in the training process.
After carrying out inference for all frequency bins, denoised images are converted to time-sequential data with inverse FT.
The segmentation image at the sound frequency is the final segmentation label since the image at this sound frequency exhibits the highest contrast between the sound field and silhouettes and is considered to be easier to separate those two classes.

\subsection{Dataset Creation}
Since no dataset exists for training the network, we created a dataset of sound fields including objects.
We used acoustic numerical simulation to create a training dataset because it is difficult to collect sound-field data under various conditions through experiments.

\begin{figure*}[t]
  \centering
   \includegraphics[width=1.0\linewidth]{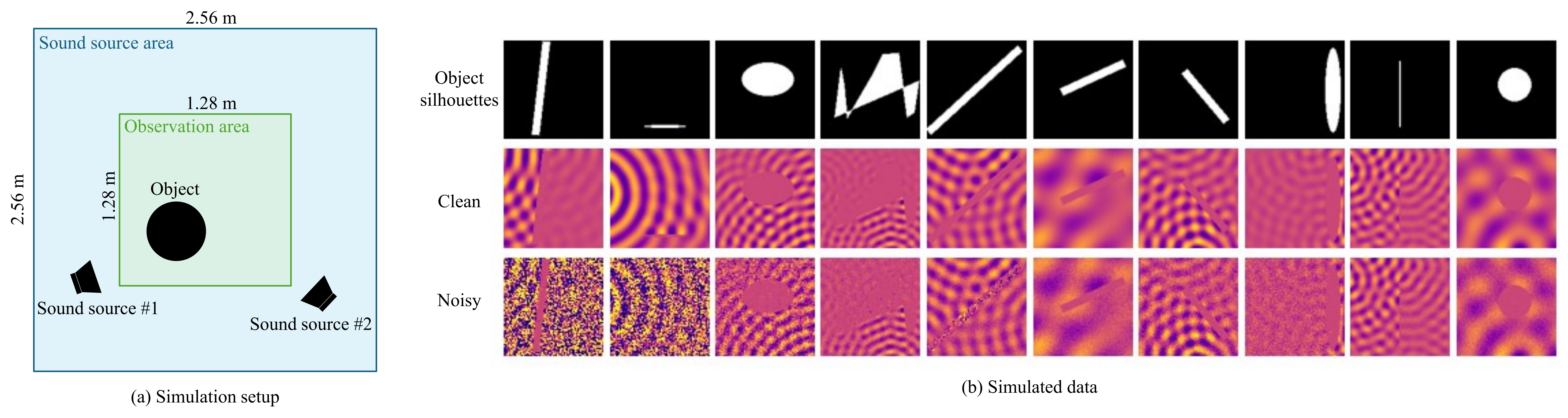}
   \caption{Dataset creation. (a) Simulation setup. Sound sources are installed outside observation area. Objects are installed inside observation area. (b) Simulated data. Top row shows object silhouettes, second row shows clean simulated images, and bottom row shows noisy images with noise added to clean images. Color indicates real part of complex amplitude, ranging from $-1.0$ to $1.0$.}
   \label{fig:dataset}
\end{figure*}

\paragraph{Simulation Conditions}
To simulate sound fields including objects, we conducted a simulation of the sound fields in the time domain with MATLAB~\cite{MATLAB} using the k-Wave toolbox~\cite{k-wave}.
The simulation setup is shown in~\cref{fig:dataset}(a).
Following DSFD, to obtain the image size of $128 \times 128$ pixels, the observation area was set to $1.28$ m $\times$ $1.28$ m, and observation points were set in a grid pattern at intervals of $0.01$ m.
Sound sources were randomly placed outside the observation area, within a range of $2.56$ m $\times$ $2.56$ m.
Objects were set inside the observation area.
The shapes of objects included ellipses, lines, and polygons, and the parameters positions were randomly selected.
The medium other than the object was air, and the object was made of expanded polystyrene (EPS) to enable stable calculations.
Thus, the reflectivity of the object was 93.2\%.

The simulation conditions are listed in~\cref{tab:simcond}.
The frequency range of sound $f_s$ was set to $90 \leq f_s \leq 2800$ Hz, which corresponds to the wavenumber $k$ being $1.66 \leq k \leq 51.7$ rad/m.
The sound frequency for each condition was fixed, which means each sound source had the same frequency and was randomly selected from a uniform distribution.
The amplitude of the first sound source was set to $1$ Pa, and those of the other sound sources were randomly selected from a uniform distribution between $0.1$ to $1$ Pa.
Calculated time-sequential data were converted to the frequency domain with FT and the data at the sound frequency were extracted.
The total amount of images was $55,000$, with $11,000$ images for each number of sound sources.
Sample simulated data are shown in the second row of \cref{fig:dataset}(b).
Reflections and diffractions were simulated to occur around the silhouettes highlighted in white in the first row of \cref{fig:dataset}(b).

\paragraph{Noisy-Data Creation}
\label{subsec:dataset}
Since the noise characteristics between sound-field regions and object-silhouette regions can be different, we added different types of noise.
For the sound-field regions, we added white noise with different SNRs.
The SNRs were randomly selected from a uniform distribution between $-20$ to $20$ dB.
For the silhouette regions, we created noise on the basis of experimentally obtained data.
The details of the noisy data creation for silhouette regions are in the supplementary material.
We estimated the probability density function (PDF) from empirical data using kernel density estimation~\cite{KDE}.
On the basis of the estimated PDF, we generated noise for silhouette regions by using the inverse transform sampling method~\cite{Inversion}.
The SNRs of the noise for silhouette regions were also selected from a uniform distribution between $-20$ to $20$ dB.
However, since clean-image values in silhouette regions are zero, SNRs were set for signals in the sound-field regions.

Created noisy data are shown in the last row of~\cref{fig:dataset}(b).
Because the SNRs of the silhouette and sound-field regions were set independently, the intensities of the noise differed.

\begin{table}
  \centering
  \begin{tabular}{c|c}
    \toprule
    Parameter & Values \\
    \toprule
    Spatial grid size [m] & 0.01 \\
    \midrule
    Temporal discretization step [s] & $1.21 \times 10^{-5}$  \\ 
    \midrule
    Speed of sound in air [m/s] & 340 \\
    \midrule
    Speed of sound in EPS [m/s] \cite{EPS2015} & 414\\ 
    \midrule
    Density of air [$\text{kg/m}^3$] & 1.21 \\
    \midrule
    Density of EPS [$\text{kg/m}^3$]\cite{EPS2015} & 28.0 \\
    \midrule
    Number of sound sources $S$ & $S \in \{ 1,2,3,4,5 \}$ \\
    \midrule
    Frequency of sound $f_s$ [Hz] & $90 \leq f_s \leq 2800$ \\
    \midrule
    Amplitude of sound source $p_s$ [Pa] & $0.10 \leq p_s \leq 1.0 $ \\
    \bottomrule
  \end{tabular}
  \vspace{-2pt}
  \caption{Simulation conditions.}
  \label{tab:simcond}
\end{table}

\begin{figure*}[t]
  \centering
   \includegraphics[width=1.0\linewidth]{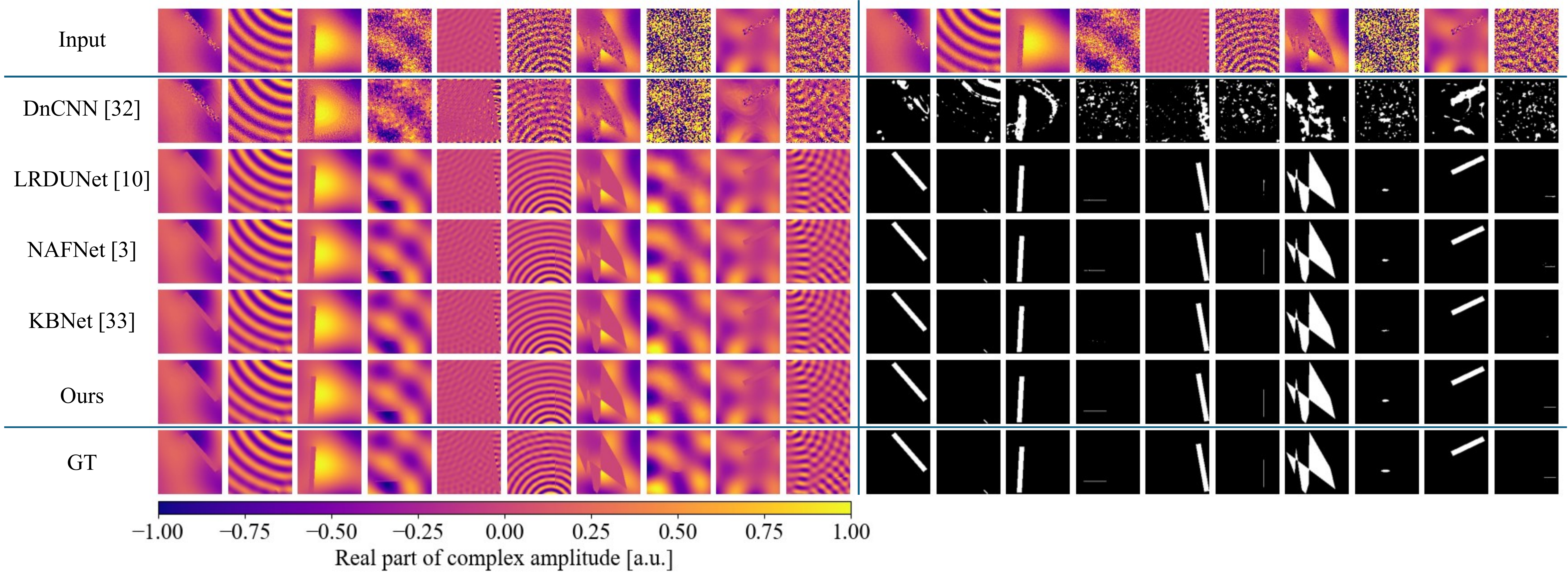}
   \caption{Qualitative results. Top row shows input images. From second to sixth rows, denoised and segmented images are shown. Last row shows GT images. Left ten columns are for denoising and right ten columns are for segmentation. Input images on right ten columns are same as those in left ten columns. Color indicates real part of complex amplitude.}
   \label{fig:result_sim}
\end{figure*}

\section{Experiment}
\label{sec:exp-sim}

\subsection{Implementation Details}
The network was implemented by PyTorch  on the basis of the official implementation of CGNet~\cite{CGNet2024, CGNet_github}.
The CGNet consists of an input layer, four encoder blocks, one middle block, four decoder blocks, and an output layer.
The numbers of encoder blocks were set to $2$, $2$, $4$, and $6$, respectively, in order of increasing depth.
The number of middle blocks was $10$.
The number of decoder blocks were set to $2$, $2$, $2$, and $2$, respectively, in order of decreasing depth.

The input image size was set to $128 \times 128$ pixels, and the number of channel was two.
The weighting coefficient $\lambda$ of the loss function in~\cref{eq:lossfun} was set to $\lambda = 10$ on the basis of experiments conducted with a small amount of data.
The $\alpha$ of the loss function in~\cref{eq:segloss} was set to $\alpha = 0.5$. 
The AdamW optimizer was used where the learning rate was 0.001, weight decay was $0.0$, and $\beta_1$ and $\beta_2$ were $0.9$ and $0.9$, respectively.
The cosine annealing scheduler was used where the maximum number of iterations was $400,000$, and the minimum learning rate was $\text{1e-7}$.
The batch size was $16$, and the number of epochs was $20$.
The training time was approximately $3.5$ hours with a single NVIDIA GeForce RTX 4090 GPU.
The number of training, validation, and evaluation images were $50,000$, $2,500$, and $2,500$, respectively.

The PSNR and structural similarity (SSIM) were used for denoising, and intersection over union (IoU) was used for segmentation as the evaluation metrics.
The IoUs were calculated for class $1$, i.e., the silhouette regions.

\subsection{Compared Models}
In our proposed method, we used CGNet as the base network.
To validate the use of CGNet, we also evaluated the performance when using existing networks as the base network.
Compared networks were selected from denoising networks since denoising is more complicated task than segmentation, which is a two class classification in our task.

The following four conventional denoising networks were used: denoising convolutional neural network (DnCNN)~\cite{DnCNN}, lightweight residual dense neural network based on the U-net neural network (LRDUNet)~\cite{LRDUNet}, NAFNet~\cite{NAFNet2022}, and kernel basis network (KBNet)~\cite{KBNet}.
Since all those four networks were for denoising, we changed the number of output channels of the last layer to be the same as that of the proposed method.
Under the training of DnCNN, LRDUNet, NAFNet, and KBNet, the $\lambda$ of the loss function in~\cref{eq:lossfun} were set to $0.001$, $0.01$, $0.005$, and $0.01$, respectively.
The batch sizes for DnCNN, LRDUNet, NAFNet, and KBNet were set to $32$, $32$, $32$, and $16$, respectively.
The number of epochs for all models was $20$.
The implementation details are in the supplementary material.

\begin{table}
  \centering
  \begin{tabular}{c|c|c|c}
    \toprule
    Base Network & PSNR [dB] & SSIM & IoU \\
    \midrule
    DnCNN~\cite{DnCNN} & 16.6 & 0.603 & 0.284 \\
    LRDUNet~\cite{LRDUNet} & 39.6 & 0.976 & 0.970 \\
    NAFNet~\cite{NAFNet2022} & 40.8 & 0.983 & 0.977 \\
    KBNet~\cite{KBNet} & 42.0 & 0.986 & 0.976 \\
    CGNet~\cite{CGNet2024} (Ours) & \textbf{43.2} & \textbf{0.987} & \textbf{0.986} \\
    \bottomrule
  \end{tabular}
  \vspace{-2pt}
  \caption{Quantitative results. Ours performed best.}
  \vspace{-4pt}
  \label{tab:result_sim}
\end{table}

\subsection{Results}
The results are shown in~\cref{tab:result_sim} and \cref{fig:result_sim}.
The table shows that our method, which used CGNet as a base network, recorded the highest scores in terms of PSNR, SSIM, and IoU.
As shown in~\cref{fig:result_sim}, both sound-field and silhouette regions were denoised except for DnCNN.
In the segmentation results, small objects that could not be detected by KBNet are successfully detected by ours.
From these results, we confirmed that using CGNet as a base network was valid for sound-field-image joint denoising and segmentation.

\begin{figure}[t]
  \centering
   \includegraphics[width=1.0\linewidth]{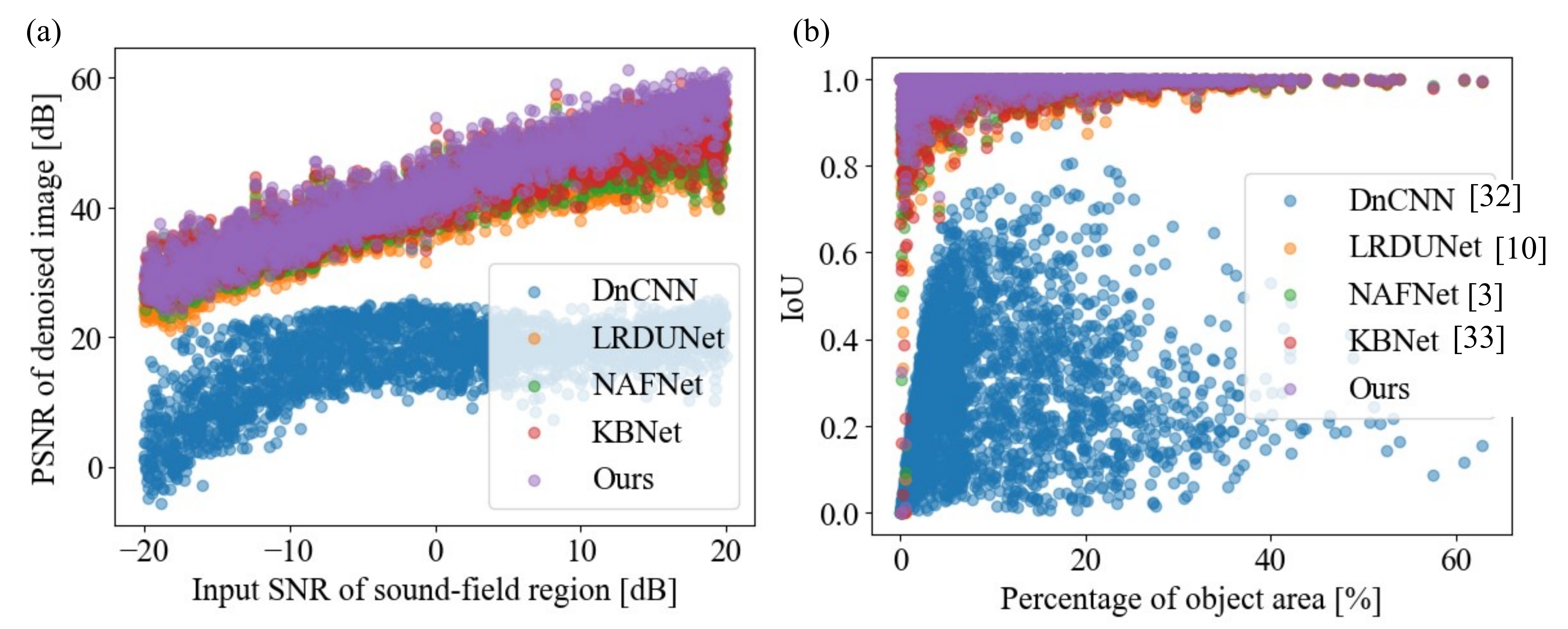}
   \caption{Analyses of results. (a) PSNRs of denoised images relative to SNRs of sound field in input images. PSNRs improved as input SNR increased except for DnCNN. (b) IoUs of segmentation images relative to percentage of object silhouettes' area. IoUs tended to decrease where areas were small.}
   \label{fig:result_sim_analysis}
\end{figure}

To analyze denoising ability, we investigated the effect of the input SNRs of sound-field regions on the PSNRs, as plotted in~\cref{fig:result_sim_analysis}(a).
The horizontal axis means the input SNRs for sound-field regions, and the vertical axis means the output PSNRs. 
Except for DnCNN (blue dots), the output PSNRs positively correlated with the input SNRs for sound-field regions.
Although the output PSNRs performed similarly where the input SNRs were low, except for DnCNN, ours (purple dots) showed improved output PSNRs at high input SNRs around $10$ to $20$ dB.

To analyze segmentation ability, we investigated the effect of the percentages of silhouette areas on the IoUs as plotted in~\cref{fig:result_sim_analysis}(b).
The percentages of silhouette areas were calculated as the number of pixels in the object area divided by the total number of pixels in the image.
Except for DnCNN (blue dots), the number of data points with low IoUs tended to increase as the size of the object area decreased.
Ours (purple dots) showed that there are many data points near 1.0 IoU even where the percentages of silhouette areas were small around $0$ to $10$\%.

\begin{figure*}[t]
  \centering
   \includegraphics[width=1.0\linewidth]{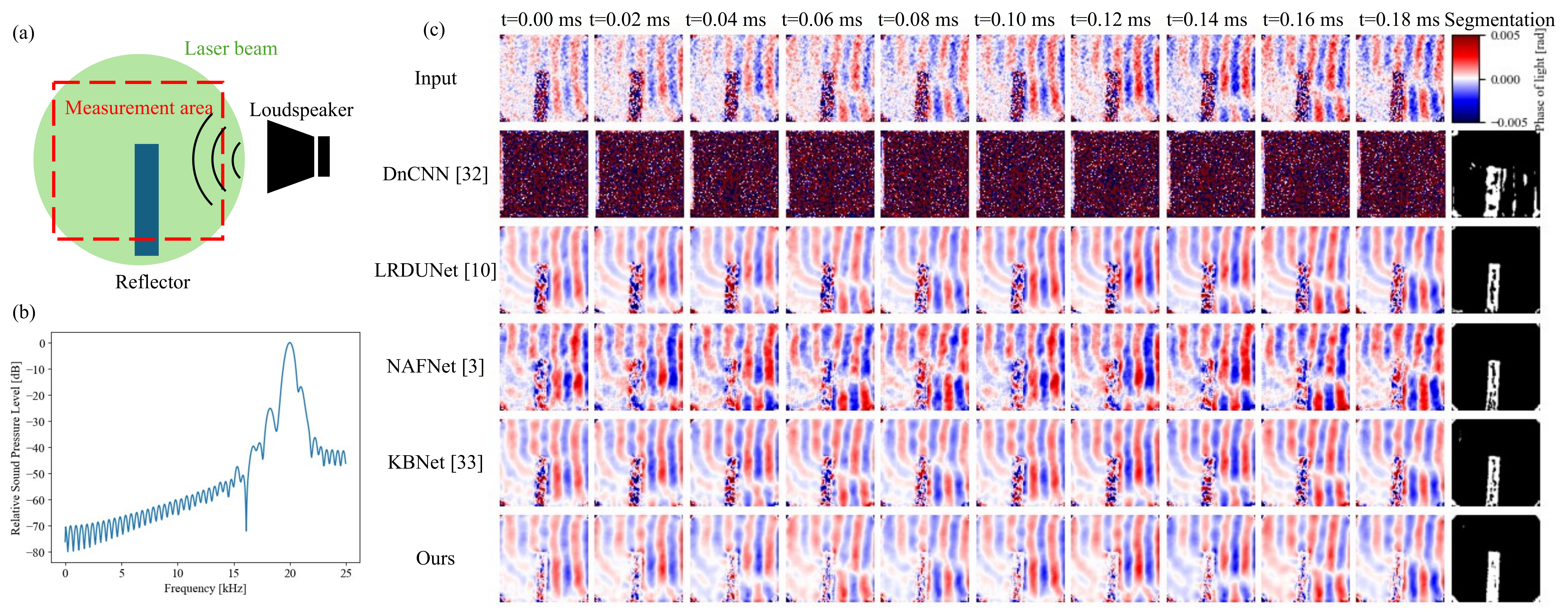}
   \caption{Experimental results of sound field with reflection and diffraction. (a) Experimental setup. Reflector was installed 15 cm from loudspeaker. (b) Frequency spectrum of input data. (c) Experimental results of denoising and segmentation. Top row shows input images in time domain and other rows show denoised and segmentation results. Color indicates phase of light detected with PPSI.}
   \label{fig:result_exp_speaker}
\end{figure*}

\begin{figure*}[t]
  \centering
   \includegraphics[width=1.0\linewidth]{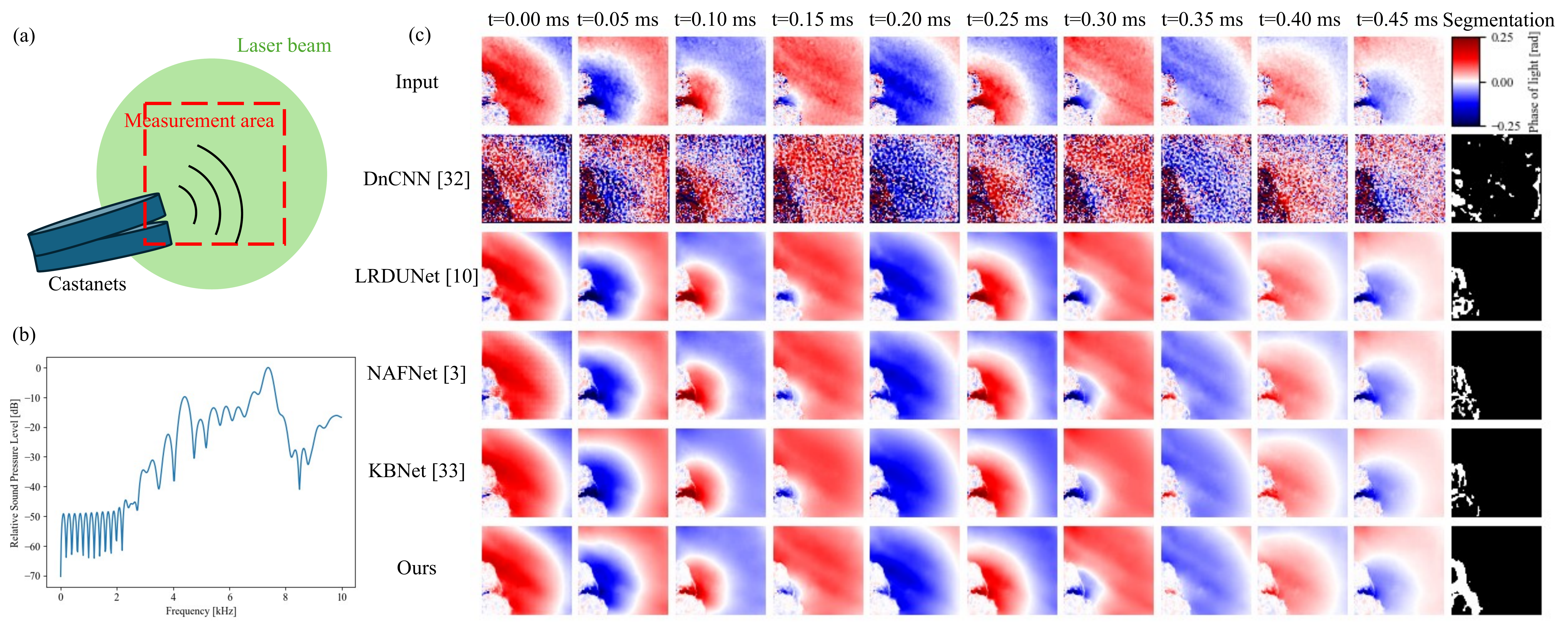}
   \caption{Experimental results of castanets sound. (a) Experimental setup. Castanets was located at bottom-left of imaging area. (b) Frequency spectrum of input data. (c) Experimental results of denoising and segmentation. Top row shows input images in time domain and other rows show denoised and segmentation results. Color indicates phase of light detected with PPSI.}
   \vspace{-4pt}
   \label{fig:result_exp_castanet}
\end{figure*}

\vspace{-2pt}
\section{Evaluation for Experimental Data}
\label{sec:exp-real}
To confirm the applicability to the experimentally measured data, we applied our method to the following two types of experimental data:
(1) A sound field diffracted with a thin plate, and (2) a sound field generated with a wooden finger castanets, a percussive musical instrument.
All the data were captured by parallel phase-shifting interferometry (PPSI)~\cite{PPSI2016}, which is often used for sound-field imaging due to its high sensitivity and spatial resolution.
PPSI can capture sound-field images within 100 mm in diameter by using a high-speed polarization camera~\cite{Onuma2014}.

\subsection{Sound Field with Reflection and Diffraction}
\paragraph{Experimental Setup}
We recorded a sound field with reflection and diffraction where a reflector was set in front of a loudspeaker.
The schematic diagram of the setup is shown in~\cref{fig:result_exp_speaker}(a).
The loudspeaker (NS-BP200, YAMAHA) was set outside the measurement area.
The reflector was set inside the measurement area and 150 mm from the loudspeaker.
The dimensions of the reflector were 150 mm in height, 10 mm in width, and 200 mm in depth.
A 20-kHz sinusoidal wave was emitted from the loudspeaker.
The frame rate of the camera was set to 50000 frames per second.
A frequency spectrum of captured data was shown in~\cref{fig:result_exp_speaker}(b).
The spectrum was calculated from a time series signal at the index of $[20, 100]$, where $[\cdot, \cdot]$ represents the pixel coordinates in terms of height and width, respectively.
Since the sound emitted from the loudspeaker was a 20-kHz sinusoidal wave, the spectrum exhibited a distinct peak at 20~kHz.

\paragraph{Results}
The results are shown in~\cref{fig:result_exp_speaker}(c).
The top row is the input data from $t=0.00$ to $t=0.18$ ms with interval of $0.02$ ms.
The sound wave propagated from right to left of the imaging area.
The reflection and diffraction occurred around the reflector.
Since the color range was kept consistent across all conditions to ensure fair visualization, the denoised results of DnCNN were saturated; therefore, it performed the worst.
Although the denoised results of LRDUNet seemed to clearly visualize the sound wave, the amplitudes of the diffracted waves seemed larger than in the input data.
Ours showed that noise was effectively removed while maintaining amplitudes of sound waves close to the input data.
Ours also excelled in denoising silhouette regions compared with the others.
In terms of segmentation, although there were undetected pixels, ours could estimate the least number of undetected pixels.

\subsection{Sound Field with Sound Source Object}
\paragraph{Experimental Setup}
We also conducted denoising and segmentation for the sound field including the sound source object.
We used wooden finger castanets, a percussive musical instrument as a sound source.
The schematic diagram of the experimental setup is shown in~\cref{fig:result_exp_castanet}(a).
The castanets was installed, the edge of which was included in the measurement area.
The castanets was played by human hand, and the sound was recorded with a PPSI system at 20,000 frames per second.
A frequency spectrum of captured data was shown in~\cref{fig:result_exp_castanet}(b).
The spectrum was calculated from a time series signal at the index of $[90, 40]$.
The spectrum exhibited a broad frequency distribution, indicating that the signal contained multiple frequencies.

\paragraph{Results}
The results are shown in~\cref{fig:result_exp_castanet}(c).
The top row is the input data from $t=0.00$ to $t=0.45$ ms with intervals of $0.05$ ms.
The sound-wave propagation from castanets, located at bottom-left corner, can be seen.
Since the color range was consistent across all conditions, the denoised results of DnCNN were saturated.
All networks except DnCNN could denoise the fine noise, and the sound wave was smoothed.
By KBNet, the wavefront's shape was more rounded, which was observed especially at $t=0.10$ and $0.30$ ms, whereas ours kept the shape of the wavefront.
Although segmentation did not work well in all models, ours could capture the edge of the castanets.
Post-processing, such as dilation, would fill the holes.

\section{Comparison of Single-tasking and Multitasking}
We evaluated how carrying out denoising and segmentation at the same time would change the accuracy compared with carrying out each as a single task.
To do so, denoising and segmentation were conducted separately on the basis of CGNet.
The results are shown in~\cref{tab:single-multi}.
The performance of denoising and segmentation by multitasking was slightly better than those of single-tasking.
We also confirmed that multitasking can be implemented with minimal impact on inference time of single image and total model size for single-image input.

\begin{figure}[t]
  \centering
   \includegraphics[width=1.0\linewidth]{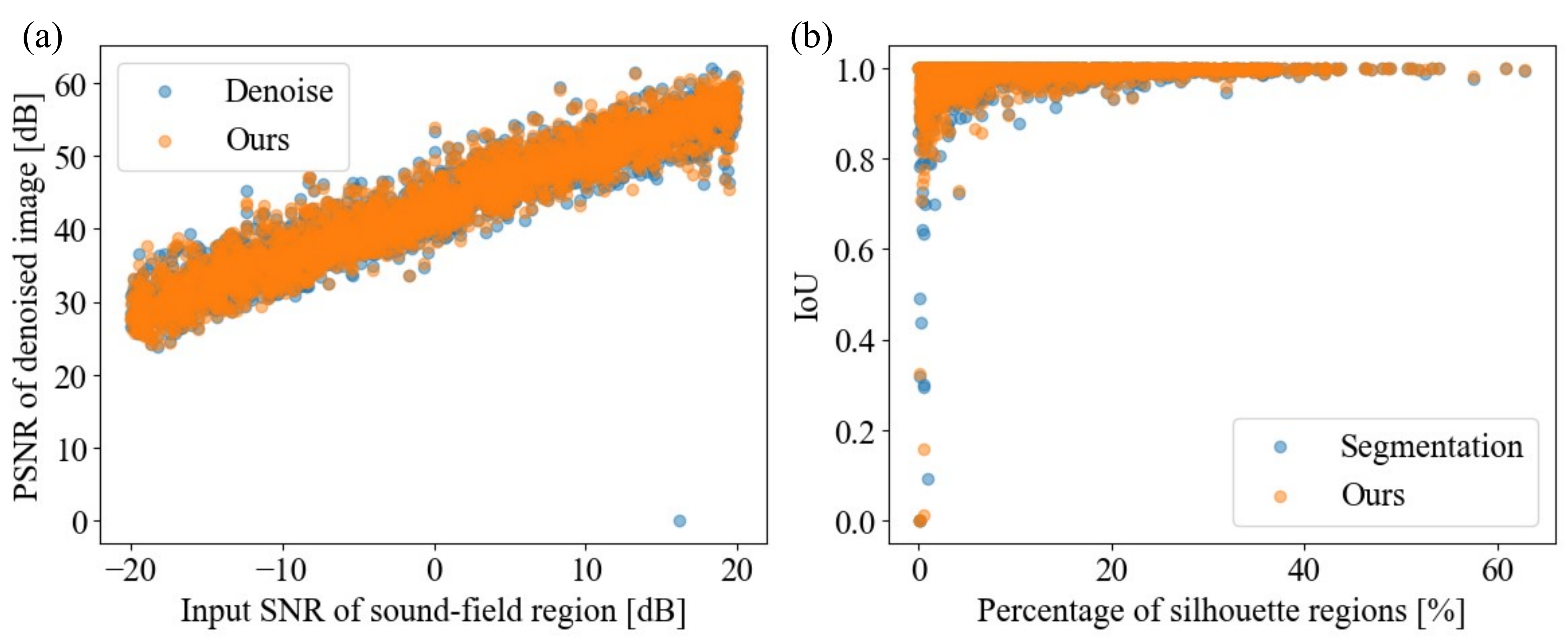}
   \caption{Comparison of single-tasking and multitasking results. (a) PSNRs of denoised images relative to input SNRs of sound region. Blue and orange dots show results of single-tasking and multitasking, respectively. No significant difference in denoising performance between single-tasking and multitasking were observed. (b) IoUs of segmentation images relative to percentage of object silhouettes' area. Multitasking IoUs (Ours) were higher where silhouette regions were small.}
   \label{fig:result_single-vs-multi}
\end{figure}

For further investigation, PSNRs relative to input SNRs and IoUs relative to the percentage of object silhouettes' region are shown in \cref{fig:result_single-vs-multi}.
From \cref{fig:result_single-vs-multi}(a), there was no significant difference in denoising performance between single-tasking and multitasking.
On the other hand, according to \cref{fig:result_single-vs-multi}(b), multitasking IoUs (blue dots) tended to be higher than single-tasking IoUs (orange dots) where object silhouettes' areas were less than 10 \%.

\begin{table}
  \centering
  \resizebox{\columnwidth}{!}{
  \begin{tabular}{c|c|c|c|c|c}
    \toprule
    \multirow{2}{*}{Task} & PSNR & \multirow{2}{*}{SSIM} & \multirow{2}{*}{IoU} & Inference & Model \\
    & [dB] & & & time [ms] & size [MB] \\
    \midrule
    Denoising & 43.1 & 0.986 & - & 21.99 & 1243.87 \\
    Segmentation & - & - & 0.984 & 21.82 & 1243.74 \\
    Ours & \textbf{43.2} & \textbf{0.987} & \textbf{0.986} & 22.10 & 1244.01 \\
    \bottomrule
  \end{tabular}
  }
  \caption{Comparison of single-tasking and multitasking. Multitasking showed slightly better performance than single-tasking without a significant increase in inference time and model size.}
  \label{tab:single-multi}
\end{table}

\section{Conclusions}
\label{sec:conclusion}
We proposed a denoising and segmentation method for 2D sound-field images with object silhouettes.
To handle the sound fields with acoustic scattering by objects, we created a dataset through acoustic simulation.
Multitasking was realized by using the output of the final layer of CGNet as channels for denoising and segmentation and calculating the loss function for each task.
We confirmed that the proposed method can be applied to both simulated and experimental data.
We believe that this method can be used for analyzing sound fields with interacting objects and for sonars in self-driving vehicles and assistive robots.
Future work includes improving segmentation performance for experimental data, evaluating with recent segmentation architectures, and enabling moving-object segmentation.

\clearpage
{\small

}

\clearpage
\beginsupplement

\section*{Supplemental Material}
\addcontentsline{toc}{section}{Supplemental Material}

We summarize the content of the supplementary material as follows. 
Section \ref{sec_supp:conventional method} presents the issue with using the existing denoising/segmentation methods in supporting the data underlying the motivation of our task.
Section \ref{sec_supp:noise creation} provides details on the creation of noise data for silhouette regions based on the experimental data.
Section \ref{sec_supp:implement} provides the implementation details of the compared models.

\section{Issue with Existing Methods}
\label{sec_supp:conventional method}

\begin{figure}[b]
  \centering
   \includegraphics[width=1.0\linewidth]{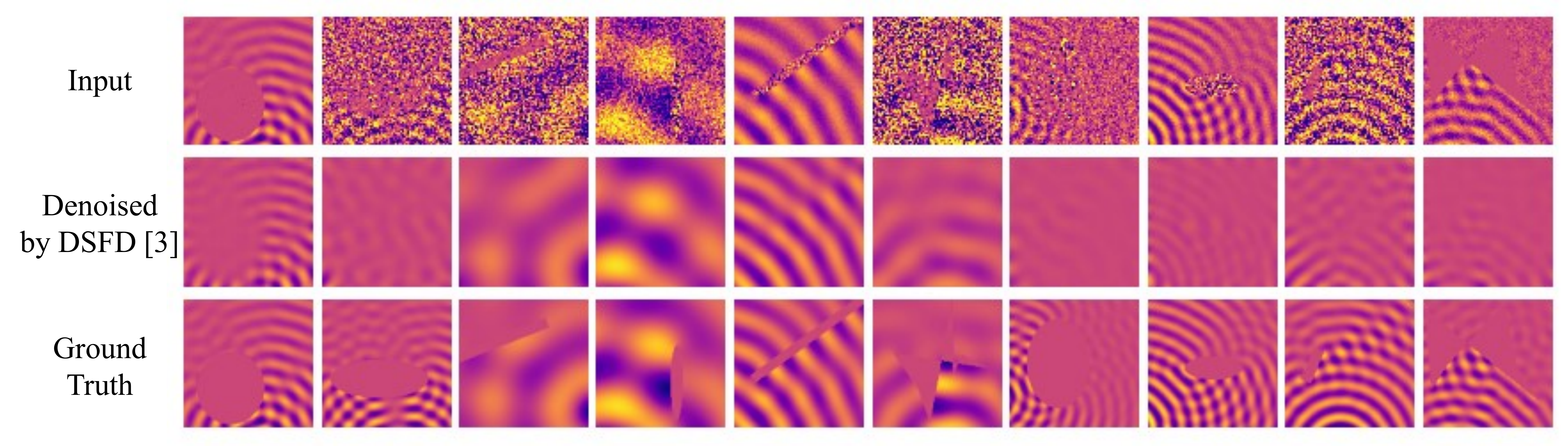}
   \caption{Denoising results estimated using DSFD trained with w/o silhouette dataset}
   \label{fig:result_woObj_wObj}
\end{figure}

DSFD~\cite{Ishikawa2023} focuses only on the sound field without objects.
Thus, it cannot be directly applied to a sound field with objects.
To provide evidence for this, denoising results for sound-field images with object silhouettes are shown in~\cref{fig:result_woObj_wObj}.
The denoising was carried out by DSFD trained with the without-silhouette (w/o silhouette) dataset.
The trained model was obtained from the publicly available GitHub repository of the author of DSFD~\cite{Ishikawa2023_github}.
We created the evaluation data, which included object silhouettes.
The sound waves appeared inside the silhouette regions on the second-row images.
Therefore, the DSFD cannot properly denoise the sound-field images, especially in the silhouette regions.

\begin{figure}[t]
  \centering
  \includegraphics[width=1.0\linewidth]{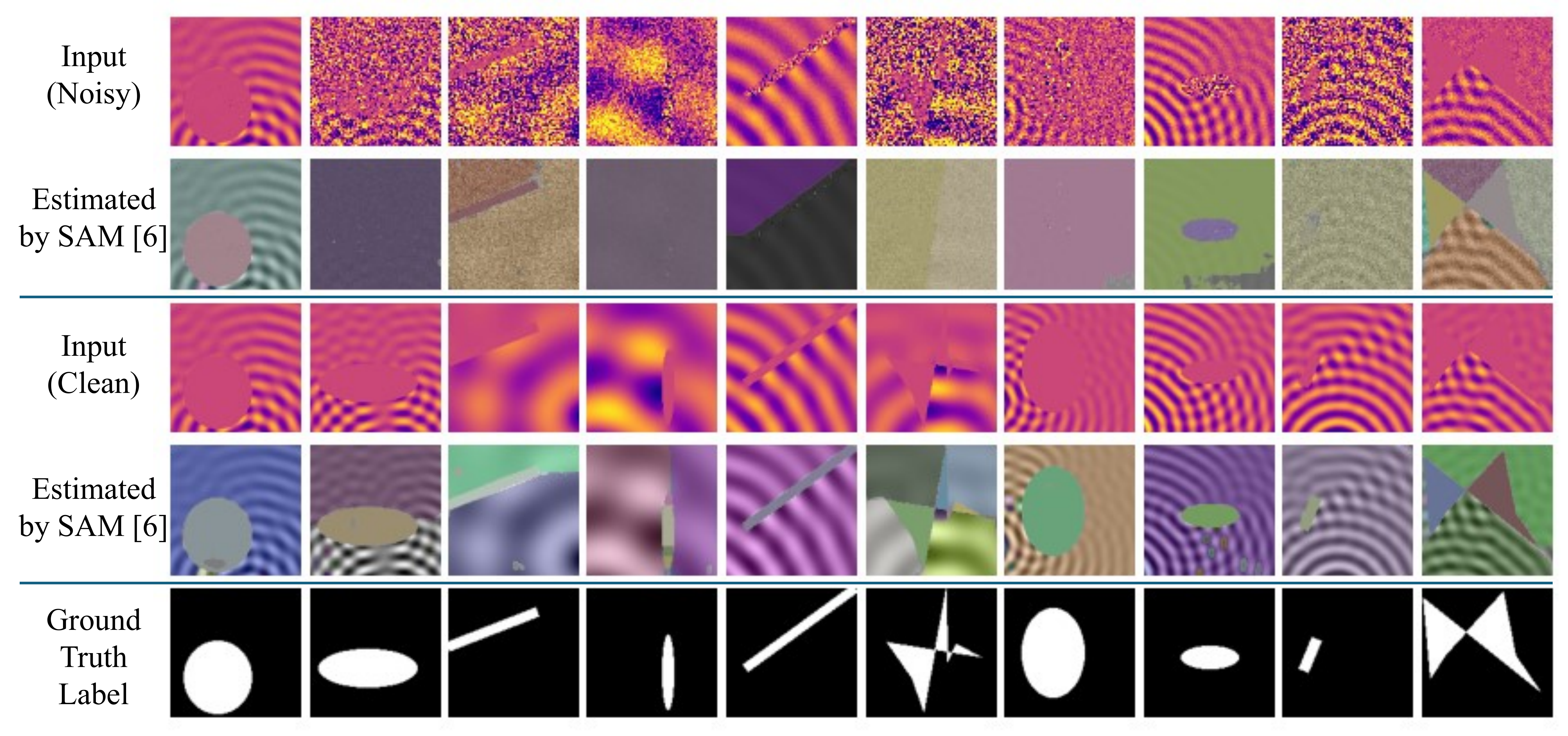}
  \caption{Segmentation results estimated by SAM}
  \label{fig:result_seg_sam}
\end{figure}

For segmentation, it may be natural to use a foundational model designed for natural image segmentation.
To confirm the applicability of the image segmentation foundation model, we conducted a preliminary experiment.
Segment Anything Model (SAM)~\cite{SAM2023ICCV} was used to estimate zero-shot segmentation labels.
To handle sound field data with SAM, we extracted only the real part channel of the input sound field images (floating-point numbers), converted them to ranging from 0 to 255, and then transformed them into 1-channel images similar to Grayscale images. 
Subsequently, these images were converted to RGB for input into SAM.
The segmentation results are shown in~\cref{fig:result_seg_sam}.
The top two rows are the input and segmented images for noisy data, the next two rows are the input and segmented images for clean data, and the last row is the ground truth of the segmentation labels.
The visualization of the segmentation masks obtained by SAM is performed by overlaying randomly assigned colors for each mask on the input images. Therefore, the same color represents a single segmentation mask.
For the results with noisy images as input (See the second-row of~\cref{fig:result_seg_sam}), the segmentation does not perform well where the noises in the input image are high, for example, the second, fourth, sixth, and ninth columns from the left in the~\cref{fig:result_seg_sam}.
For the results with clean images as input (See the fourth-row of~\cref{fig:result_seg_sam}), there are no images where the object silhouettes are entirely unsegmented.
However, some images show multiple segments within the same object silhouettes, for example, the first, second, and fourth columns from the left in the~\cref{fig:result_seg_sam}.
From these results, it can be concluded that even with the foundation model for image segmentation, SAM, attempting zero-shot segmentation on the noisy data is ineffective. 
Furthermore, the performance, even with the clean data, is inadequate.
Hence, we considered the task of joint training and inferring denoising and segmentation.

\section{Noise Creation based on Experimental Data}
\label{sec_supp:noise creation}
As mentioned in the main paper regarding dataset creation, we calculated the noise for silhouette regions from experimentally obtained data.
In this section, we provide supplemental information for data collection and noise creation.

\begin{figure}[t]
  \centering
   \includegraphics[width=1.0\linewidth]{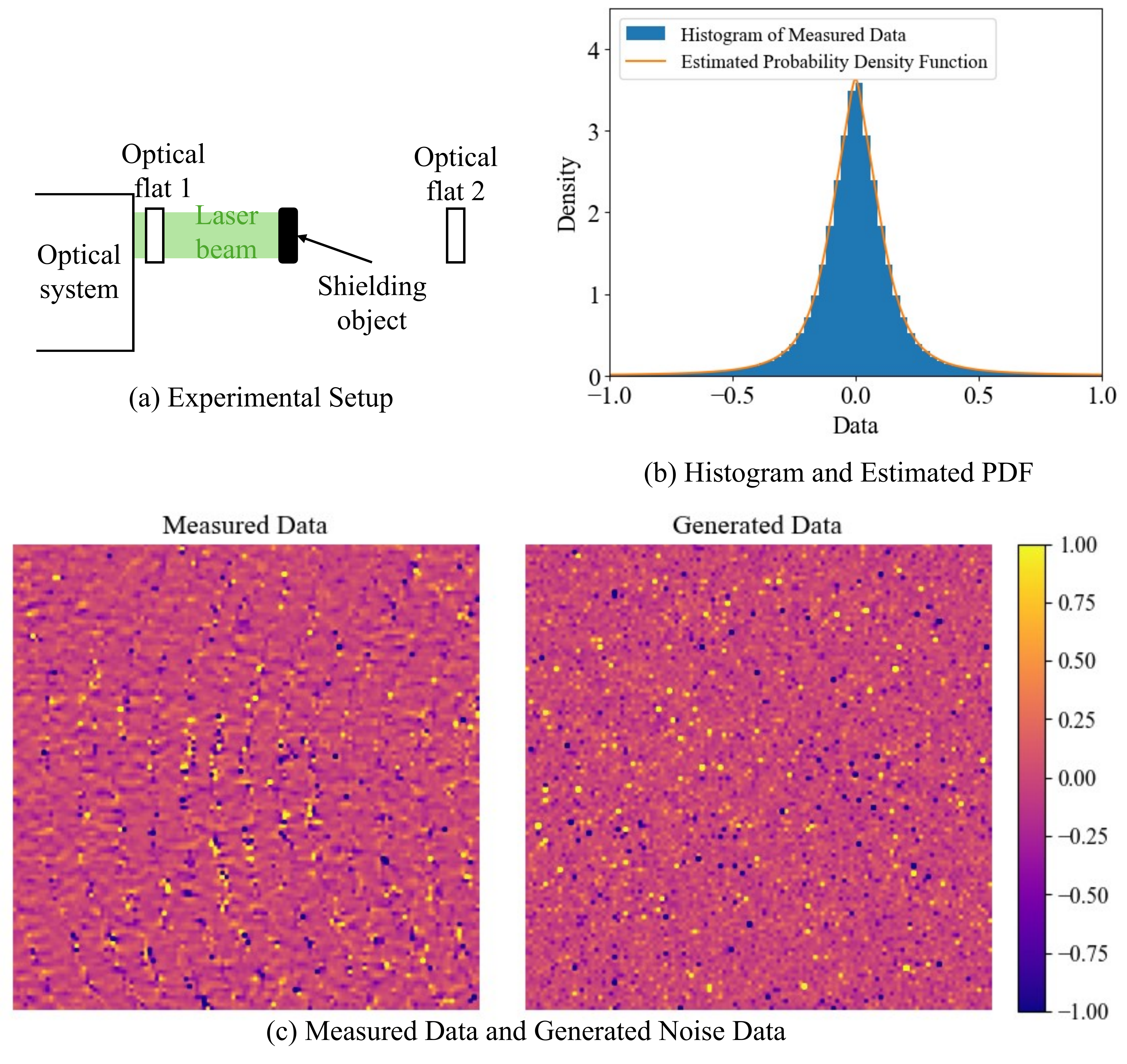}
   \caption{Noise-data creation based on experimental data. (a) Experimental setup of data collection. (b) Histogram of measured data and estimated probability density function (PDF). (C) Measured data and generated noise data.}
   \label{fig:created noise}
\end{figure}

We estimated the probability density function (PDF) by kernel density estimation (KDE) on the basis of experimentally measured data.
The data were collected by parallel phase-shifting interferometry (PPSI)~\cite{PPSI2016}.
The experimental setup is shown in~\cref{fig:created noise}(a).
We installed a shielding object between two optical flats and recorded the data five times.
The frame rate of the high-speed camera in the PPSI system was set to 20,000 frames per second, and 200 images were collected for each recording.
To remove the low-frequency noise, a high-pass filter with a 500-Hz cut-off frequency was applied to the recorded images along the time direction.
The real and imaginary parts of the Fourier-transformed data were regarded as one image.
The single pixel value was regarded as one sample, and 28,800,000 samples in total were used for estimating the PDF.
Histograms of the measured data and estimated PDF are shown in~\cref{fig:created noise}(b).
There is good agreement between the estimated PDF and histogram of the measured data.

Noise data for silhouette regions were generated on the basis of the estimated PDF by using the inverse transform sampling method.
An example of the generated data is shown in~\cref{fig:created noise}(c).
The left and right figures show the measured and generated data, respectively.
The generated data were sampled data based on the estimated PDF corresponding to the number of pixels in the image and reshaped to match the image dimensions.
We confirmed that the generated noise data was similar to the measured data, except for spatial patterns originating from the optical elements.

\section{Preliminary experiment for loss function}
\label{sec_supp:lossfun}
To determine the loss function for the proposed method, we conducted a preliminary experiment to compare performance by loss functions.
For denoising loss $L_{\rm denoise}$, mean squared error (MSE), mean absolute error (L1), and negative peak signal-to-noise ratio (N-PSNR) losses were compared.
For segmentation loss $L_{\rm seg}$, binary cross entropy (BCE) and balanced BCE and dice (BCE+Dice) losses were compared.
We conducted training and evaluation with 6 patterns of all combinations of 3 loss functions for denoising and 2 loss functions for segmentation.
The evaluation result is shown in~\cref{tab:lossfun}.
The weighting coefficient $\lambda$ was set to roughly matching digits of loss values. 
Using N-PSNR as $L_{\rm denoise}$ was the best performance for denoising.
For segmentation, using BCE+Dice loss as $L_{\rm seg}$ was the best performance for segmentation.
Since combination of N-PSNR and BCE+Dice marked best performance in both denoising and segmentation, we selected them as loss functions for proposed method.

\begin{table}
  \centering
  \resizebox{\columnwidth}{!}{
  \begin{tabular}{c|c|c|c|c|c}
    \toprule
    $L_{\rm denoise}$ & $L_{\rm seg}$ & $\lambda$ & PSNR [dB] & SSIM & IoU \\
    \midrule
    MSE & BCE & $0.001$ & 40.8 & 0.983 & 0.981 \\
    MSE & BCE+Dice & $0.001$ & 41.5 & 0.984 & 0.984 \\
    L1 & BCE & $0.01$ & 42.2 & 0.986 & 0.980 \\
    L1 & BCE+Dice & $0.01$ & 42.3 & {\bf 0.987} & 0.982 \\
    N-PSNR & BCE & $10$ & {\bf 43.2} & {\bf 0.987} & 0.985 \\
    N-PSNR & BCE+Dice & $10$ & {\bf 43.2} & {\bf 0.987} & {\bf 0.986} \\
    \bottomrule
  \end{tabular}
  }
  \caption{Comparsion of loss function. Negative PSNR loss and balanced BCE and Dice loss were best for denoising and segmentation, respectively.}
  \label{tab:lossfun}
\end{table}

\section{Implementation Details}
\label{sec_supp:implement}

In this section, the implementation details for the compared models are provided.
The following parameters were common to all models.
All models were implemented by PyTorch.
The loss function for segmentation $L_{\rm{seg}}$ was the combination of binary cross entropy loss $L_{\rm{BCE}}$ and dice loss $L_{\rm{Dice}}$: 
$L_{\rm{seg}} = (1-\alpha) L_{\rm{BCE}} + \alpha L_{\rm{Dice}}$ with the weighting coefficient $\alpha=0.5$.
The number of epochs was set to $20$.
The number of channels of input and output layers were set to $2$ and $3$, respectively.
The parameters that differ for each model are listed below.

\paragraph{DnCNN~\cite{DnCNN}}
The denoising and segmentation model based on DnCNN was implemented by referencing publicly available code from the DSFD repository~\cite{Ishikawa2023_github}.
The network architecture was almost the same as in the original paper~\cite{DnCNN} except for the number of input/output channels.
The Adam optimizer was used where the learning rate was $0.001$, and $\beta_1$ and $\beta_2$ were $0.9$ and $0.999$, respectively.
The exponential learning rate scheduler was used where the multiplicative factor $\gamma$ was $0.95$.
MSE loss was used as the loss function for denoising $L_{\rm{denoise}}$.

\paragraph{LRDUNet~\cite{LRDUNet}}
The denoising and segmentation model based on LRDUNet was implemented by referencing publicly available code from the DSFD repository~\cite{Ishikawa2023_github}.
The network architecture was almost the same as in the original paper~\cite{LRDUNet} except for the number of input/output channels.
The Adam optimizer was used where the learning rate was $0.001$, and $\beta_1$ and $\beta_2$ were $0.9$ and $0.999$, respectively.
The exponential learning rate scheduler was used where the multiplicative factor $\gamma$ was $0.95$.
L1 loss was used as the loss function for denoising $L_{\rm{denoise}}$.

\paragraph{NAFNet~\cite{NAFNet2022}}
The denoising and segmentation model based on NAFNet was implemented by referencing publicly available code from the DSFD repository~\cite{Ishikawa2023_github}.
The network architecture was almost the same as in the original paper~\cite{NAFNet2022} except for the number of input/output channels.
The Adam optimizer was used where the learning rate was $0.001$, and the $\beta_1$ and $\beta_2$ were $0.9$ and $0.999$, respectively.
The exponential learning rate scheduler was used where the multiplicative factor $\gamma$ was $0.95$.
MSE loss was used as the loss function for denoising $L_{\rm{denoise}}$.

\paragraph{KBNet~\cite{KBNet}}
The denoising and segmentation model based on KBNet was implemented by referencing publicly available code from the KBNet repository~\cite{KBNet_github}.
The network architecture was almost the same as in the original paper~\cite{KBNet} except for the number of input/output channels.
The AdamW optimizer was used where the learning rate was $\text{3e-4}$, weight decay was $\text{1e-4}$, and $\beta_1$ and $\beta_2$ were $0.9$ and $0.999$, respectively.
The cosine annealing with the restart learning rate scheme was used where the periods for each cosine annealing cycle were set to $92000$ and $208000$, the restart weights at each restart iteration were all set to $1$, and the minimum learning rates at each cycle were set to $\text{3e-4}$ and $\text{1e-6}$.
L1 loss was used as the loss function for denoising $L_{\rm{denoise}}$.

\section{Evaluation of Denoising Performance for Sound Fields without Objects}
To confirm the applicability of the proposed method to sound-field images without object silhouettes, we created an evaluation dataset without objects.
The parameters of the dataset, such as the positions, frequencies, and sound pressures of the sound sources, are the same as those of the dataset described in Sec 3.2 of the main paper except for the existence of objects.
The evaluation result is shown in~\cref{tab:result-woobj}.
The trained model with the with-silhouette dataset (w/ silhouettes) was used for the evaluation.
The IoU was calculated for class 0 (sound fields), where w/o silhouette data was used for the evaluation.
These results indicate that the proposed method can be applied to sound fields without objects even if the network is only trained on data w/ silhouettes.

\begin{table}
  \centering
  \resizebox{\columnwidth}{!}{
  \begin{tabular}{c|c|c|c}
    \toprule
    Evaluation data & PSNR [dB] & SSIM & IoU \\
    \midrule
    w/o silhouettes & 43.5 & 0.991 & 1.00 (for class 0) \\
    w/ silhouettes & 43.2 & 0.987 & 0.986 (for class 1) \\
    \bottomrule
  \end{tabular}
  }
  \caption{Evaluation for w/o silhouette dataset}
  \label{tab:result-woobj}
\end{table}

\begin{figure}[t]
  \centering
   \includegraphics[width=1.0\linewidth]{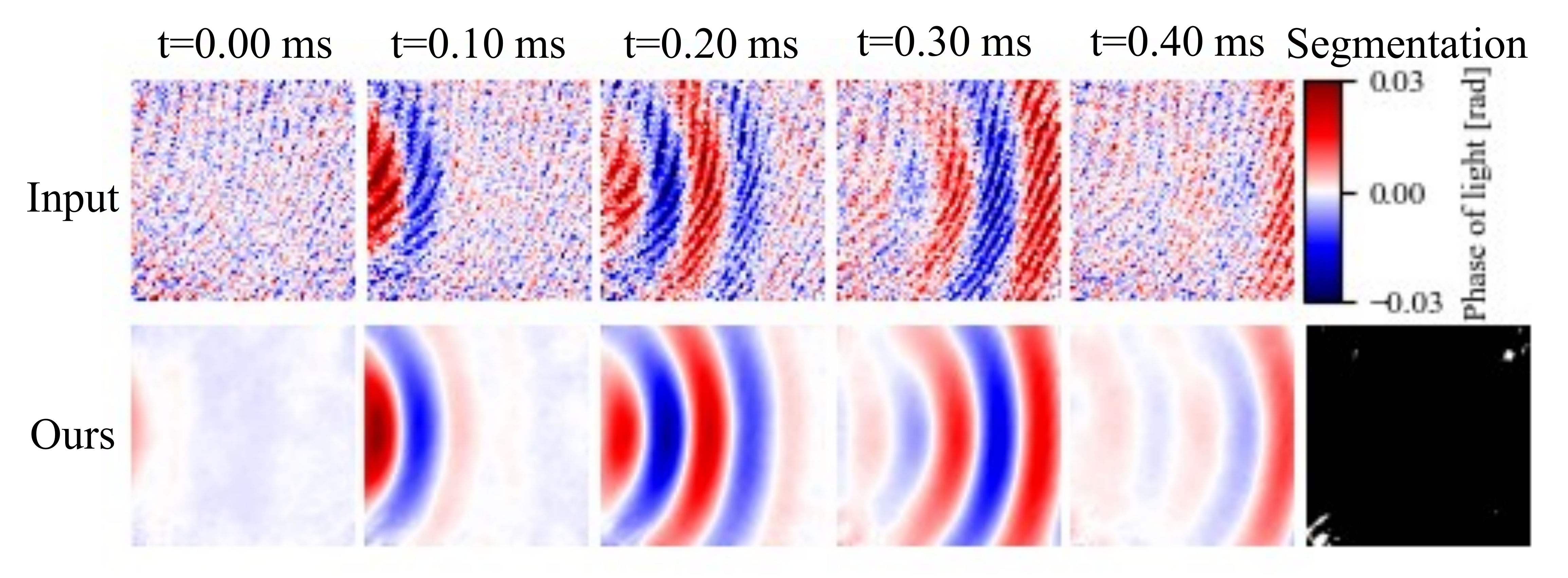}
   \caption{Experimental results of w/o silhouette sound field. Color indicates phase of light detected with PPSI.}
   \label{fig:result_exp_burst}
\end{figure}

For further verification, the results applied to the experimental data without objects are shown in~\cref{fig:result_exp_burst}.
In this experiment, sound images of a 12-kHz burst wave generated from a loudspeaker (FOSTEX FT48D)~\cite{Ishikawa2023} were used.
The top row is the input data where the burst wave propagated from left to right.
The noise was eliminated by our method.
For segmentation, although all values should be 0 (black), some pixels were falsely detected as silhouette class (white).

{\small

}

\end{document}